\documentclass[12pt]{article}
\usepackage{epsfig}
\usepackage{amsmath}
\usepackage{amssymb}
\usepackage[square,sort&compress]{natbib}

\def\slash#1{\setbox0=\hbox{$#1$}#1\hskip-\wd0\hbox to\wd0{\hss\sl/\/\hss}}

\begin{document}
\begin{titlepage}
\pagestyle{empty}
\begin{flushright}
        MZ-TH/05-13\\
\end{flushright}
\vskip 0.7in
\begin{center}
{\large {\bf Azimuthal correlation between the $(\vec{p}_\ell,\vec{p}_{X_b})$
and $(\vec{p}_\ell,\vec{P}_t)$ planes \\ in the semileptonic rest frame decay of a polarized 
top quark: \\ 
An $O(\alpha_s)$ effect}}
\end{center}
\begin{center}
 
\vskip 0.1in
{\large \bf S.~Groote$^{1,2}$, W.S.~Huo$^{1,3}$, A.~Kadeer$^1$ and  
J.G.~K\"orner$^1$} 
\\[.4truecm]

$^1$Institut f\"ur Physik der Johannes-Gutenberg-Universit\"at,\\
  Staudinger Weg 7, D-55099 Mainz, Germany\\[.3truecm]
$^2$Tartu \"Ulikooli Teoreetilise F\"u\"usika Instituut,\\
  T\"ahe 4, EE-51010 Tartu, Estonia\\[.3truecm]
$^3$Department of Physics, Xinjiang University,\\
  Shengli Road 14, 830046 \"Ur\"umqi, P.~R.~China

\vskip 0.2in {\bf Abstract}
\end{center}
\baselineskip=18pt 
\noindent
The azimuthal correlation between the planes formed by the vectors
 $(\vec{p}_\ell,\vec{p}_{X_b})$
and $(\vec{p}_\ell,\vec{P}_t)$ in the semileptonic rest frame decay of a 
polarized
top quark $t(\uparrow) \to X_b + l^+ + \nu_\ell$ belongs to a class of
polarization observables involving the top quark which vanish at the
Born term level in the standard model. We determine the next--to--leading 
order QCD corrections
to the afore-mentioned azimuthal correlation and compare the result to the 
corresponding contribution of a non--standard--model right--chiral quark current.  

\end{titlepage}

\section{Introduction}
The azimuthal correlation between the $(\vec{p}_{\ell},\vec{p}_{X_b})$
and $(\vec{p}_{\ell},\vec{P}_t)$ planes in the semileptonic rest frame decay of a 
polarized top quark (see Fig.~\ref{Cazi}) belongs to a class of polarization observables
involving the top quark in which the leading--order (LO) contribution gives a
zero result in the Standard Model (SM). As we shall see later on, the
vanishing of this azimuthal correlation is a 
consequence of the
left--chiral $(V-A)(V-A)$ nature of the current--current interaction in the SM. 
Another example of a LO zero polarisation observable is the decay of a top quark 
into a polarized 
transverse-plus $W$ boson and a (massless) bottom quark where the rate
into the transverse-plus $W$ boson is zero at the Born term level due to the
left--chiral (V-A) coupling structure of the SM. Still 
another example is the production of longitudinally polarized top quarks in 
$e^+-e^-$ annihilation produced from the longitudinal part of the
intermediate gauge bosons (Z and/or $\gamma$). The corresponding rate is
zero due to the absence of second--class currents in the SM.

For the latter two above cases 
the next--to--leading--order (NLO) corrections have been computed
in \cite{gkt95} and \cite{dgkm02}. In \cite{gkt95} we determined the
NLO QCD corrections to longitudinally polarized top quarks from the
longitudinal part of the intermediate gauge bosons (Z and/or $\gamma$)
in $e^+-e^-$--annihilation. The NLO QCD and
electroweak corrections to transverse-plus $W$ bosons in top quark decays
have been calculated in \cite{dgkm02}. The purpose of this note is to 
determine the
NLO QCD corrections to the afore-mentioned azimuthal correlation in polarized
top quark decay. We compare the results with the 
corresponding contribution of a non--SM right--chiral quark current. 

Nonzero contributions to the afore--mentioned polarization observables can 
either arise from non--SM effects or from higher order SM radiative 
corrections. Clearly it is important to determine the size of the NLO corrections 
to the afore--mentioned polarization 
observables before non--SM effects can be claimed to be responsible for 
nonzero values of these polarization observables.

We mention that highly polarized top quarks will become
available in singly produced top quarks at hadron colliders 
(see e.g.\cite{Mahlon:1996pn}) and in top quark pairs produced in future 
linear $e^+-e^-$ colliders (see e.g. \cite{Kuhn:1983ix,Kuhn:1985ps,
Korner:1993dy,Groote:1995ky, Groote:1996nc,Parke:1996pr,Brandenburg:1998xw}). 
It will then be possible to experimentally measure the azimuthal 
correlation
between the  $(\vec{p}_{\ell},\vec{p}_{X_b})$ and $(\vec{p}_{\ell},\vec{P}_t)$ planes.
To define the planes one needs to measure the momentum directions of the 
momenta $\vec{p}_{\ell}$ and $\vec{p}_{X_b}$ and the polarization direction of 
the top quark. The momentum direction of $\vec{p}_{\ell}$ can be directly
measured, whereas the measurement of the momentum direction of 
$\vec{p}_{X_b}$ requires the use of a jet finding algorithm. The direction
of the polarization of the top quark must be obtained from theoretical
input. In $e^+-e^-$ interactions the degree of polarization of the top quark
can be tuned with the help of polarized beams \cite{Parke:1996pr}. For 
sufficiently high energies the polarization of the top quark will be 
longitudinal in both production processes, i.e. it will point in the 
direction of its motion.
The measurement or a bound on the afore-mentioned azimuthal correlation in
polarized top quark decays will be difficult, but may yet be feasible as the
recent measurements of the helicity content of the $W$ boson in semileptonic top
quark decays by the CDF and D0 collaborations have shown \cite{W-hel-CDF-07,W-hel-DO-07}.

\section{Angular rate structure}
We shall closely follow the notation of \cite{kp99} where D.~Pirjol and one 
of us discussed the inclusive semileptonic rest frame decay of a polarized bottom 
baryon $\Lambda_b$. Of course one needs to take into account the necessary
modifications when going from the $(b \to c)$-- to the $(t \to b)$--case.
Ref.~\cite{kp99}
also contains a discussion of nonperturbative effects in the inclusive decay of
the polarized $\Lambda_b$ which were treated in next--to--leading order of 
heavy quark effective theory (HQET). This
is not necessary in the present application since the top quark decays 
essentially as a free quark.

\begin{figure}[tbhp]
\begin{center}
\includegraphics[height=4cm]{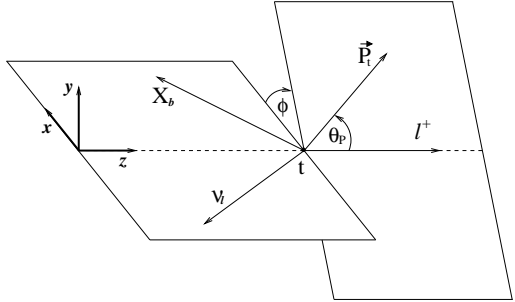}
\caption{\small \sf The definition of the azimuthal angle $\phi$ and the polar angle 
$\theta_P$ in the rest frame decay of a polarized top quark. The event plane
defines the $(x,z)$ plane. The momenta of the top quark, bottom quark, charged lepton
and the neutrino are denoted by $p_{t}$, $p_{b}$, $p_{\ell}$ and $p_{\nu}$. 
$\vec{P}_{t}$ is the polarization vector of the top quark.}
\label{Cazi}
\end{center}
\end{figure}

The general angular decay distribution of a polarized top quark decaying 
into a jet $X_b$ with bottom quantum numbers and a charged lepton $\ell^+$ and a 
neutrino $\nu_{\ell}$ is given by \cite{kp99} 
\begin{align}
\frac{d\Gamma}{dx d\cos\theta_P d\phi}=
\frac{1}{4\pi}\Big(\frac{d\Gamma_A}{dx} +\frac{d\Gamma_B}{dx}\,P\cos\theta_P
 +\frac{d\Gamma_C}{dx}\, P\sin\theta_P\cos\phi \Big)\,
\end{align}
where the polar and azimuthal angles $\theta_P$ and $\phi$ are defined in
Fig.~\ref{Cazi}. In the classification of \cite{kp99} this is the system 1b where 
the $z$--axis is determined by the lepton's momentum and $\vec{p}_{X_b}$
has a positive $x$--component. As usual we have defined a scaled lepton
energy through $x=2E_{\ell}/m_t$. 
%
%
$P$ is the magnitude of the top quark polarization. 
$d\Gamma_A/dx$ corresponds to the unpolarized differential 
rate. $d\Gamma_B/dx$ and $d\Gamma_C/dx$ describe the 
polar and azimuthal correlation between the polarization of the top quark 
and its decay products, respectively. 

The radiative corrections to the rate $\Gamma_A$ \cite{jk89} and the 
polar correlation
function $\Gamma_B$ \cite{cjk91,cjkk94,cj94} have been studied extensively 
before. We have
repeated the calculations and have found agreement with the results
in \cite{jk89,cjk91,cjkk94,cj94}. The radiative corrections to the azimuthal 
correlation function $\Gamma_C$
have not been done before. As we shall explicitly see in the next section
the LO Born term contribution to $\Gamma_C$ vanishes as was mentioned.
Technically this means that one does not have to introduce any
IR regularization scheme such as a fictituous gluon mass or dimensional 
regularization when calculating the azimuthal correlation
since at NLO the virtual one-loop and the tree--graph (real emission)
contributions are separately infrared (IR) finite. 

\section{Born term results}

It is straightforward to calculate the Born term contribution to the
decay $t(\uparrow) \to X_b + \ell^+ + \nu_{\ell}$. In the narrow resonance
approximation for the $W^+$ boson the differential rates are given by
($x=2E_{\ell}/m_t$)
\begin{eqnarray}
\label{sm-bornterm1}
\frac{d\Gamma_A^{(0)}}{dx}&=&\frac{d\Gamma_B^{(0)}}{dx}=\Gamma_F \, 
2\pi\frac{m_W}{\Gamma_W}6x(1-x)y^2 , \\
\label{sm-bornterm2}
\frac{d\Gamma_C^{(0)}}{dx}&=&0 \, , 
\end{eqnarray} 
where 
\begin{equation}
\label{rateAB}
\Gamma_F=\frac{G_F^2 m_t^5}{192 \pi^3} |V_{tb}|^2
\end{equation}
is a reference rate corresponding to a (hypothetical) pointlike four-Fermion 
interaction and
$y^2=m_W^2/m_t^2$. Note that we put the bottom quark mass to zero throughout the paper 
except for Sec.~5 where we discuss non--SM effects. 

For the 
integrated rates we obtain ($y^2\leqslant x \leqslant 1$)
\begin{eqnarray}
\Gamma_A^{(0)}&=&\Gamma_B^{(0)}=\Gamma_F \, 2\pi\frac{m_W}{\Gamma_W}
y^2(1-y^2)^2(1+2y^2) , \\
\Gamma_C^{(0)}&=&0 \, . 
\end{eqnarray}
One can read off from (\ref{rateAB}) that the width of the top quark is enhanced 
by a factor of
$2\pi m_W/\Gamma_W \cdot y^2(1-y^2)^2(1+2y^2)=44.09$ compared to a point-like 
four-Fermion interaction due to the presence of the $W$--pole 
($\Gamma_{W}=2.141~{\rm GeV}, m_{W}=80.403~{\rm GeV}$).

Let us return to Eq.~(\ref{sm-bornterm1}). The fact that $\Gamma_A=\Gamma_B$ 
means that the proposed polar correlation
measurement has 100\% analyzing power to analyze the polarization of the 
top quark whereas the azimuthal correlation measurement has zero analyzing
power. In the following we shall present some simple arguments to show
 that $\Gamma_A=\Gamma_B$ can be directly traced 
to the fact that we are dealing with a $(V-A)(V-A)$ current--current 
structure in this transition. Once this is established we then present
a physics argument that $\Gamma_C=0$ necessarily follows.

Let us rewrite the original $(V-A)(V-A)$ SM form into a more convenient
form using the Fierz transformation of the second kind which transforms
the $(V-A)(V-A)$ form into a $(S+P)(S-P)$ form (see e.g.~\cite{p83}):
\begin{align}
\label{fierz1}
M &=
\bar{u}(b)\gamma^\mu(1-\gamma_5)u(t)\,\,\bar{u}(\nu)\gamma_\mu(1-\gamma_5)v(\ell)
\\[1mm]
\label{fierz2}
&=2\bar{u}(b)(1+\gamma_5)C\bar{u}^T(\nu)\,\,v^T(\ell)C^{-1}
(1-\gamma_5)u(t)\\[1mm]
\label{fierz3}
&=2\bar{u}(b)(1+\gamma_5)v(\nu)\,\,\bar{u}(\ell)
(1-\gamma_5)u(t)
\end{align}
\noindent where we have used $C\bar{u}^T(\nu)=v(\nu)$ and 
$v^T(\ell)C^{-1}=\bar{u}(\ell)$.
The advantage of the form of Eq.~(\ref{fierz3}) is that the spinors of the top quark and the
lepton are now connected by one Dirac string. In particular this means that
there is no correlation between the top quark spin and the momenta of the
b-quark jet or the neutrino, i.e. there will be no azimuthal correlation term.
Returning to the spinor amplitude $\bar{u}(\ell)(1-\gamma_5)u(t)$ one notes that the 
combination 
$(1-\gamma_5)$ acts to project out the positive helicity spinor of the 
(massless) lepton. One can evaluate the amplitude $\bar{u}(\ell)(1-\gamma_5)u(t)$ 
for a top quark polarized in the $(\theta_P,\phi)$--direction (see Fig.~\ref{Cazi}) 
using $u_+(t)^T=\sqrt{2m_t}
( \cos\theta_P/2, e^{i\phi}\sin\theta_P/2, 0,0) $
and $\bar{u}_+(\ell)=\sqrt{E_{\ell}}(1,0,
-1,0)$ for a positive helicity lepton moving in the $z$--direction. One obtains
\begin{equation}
\label{hel-amp}
\bar{u}_+(\ell)(1-\gamma_5)u_+(t)=2\sqrt{2E_{\ell}m_t}\cos\frac{\theta_P}{2}\, .
\end{equation}
On squaring the amplitude in Eq.~(\ref{hel-amp}) one finally obtains
\begin{equation}
|\bar{u}_+(\ell)(1-\gamma_5)u_+(t)|^2=
4E_{\ell}m_t(1+\cos\theta_P)\, .
\end{equation}
An identical result is of course obtained by evaluating the trace
\begin{align}
\label{trace}
\sum_{s_{b},s_{\ell},s_{\nu}}|M|^2 &=4 \rm{Tr}\big( \slash{p}_{b}(1+\gamma_5) \slash{p}_{\nu}
(1-\gamma_5)\big)\rm{Tr}\left( \slash{p}_{\ell} (1-\gamma_5)
(\slash{p}_t
+m_t){\textstyle \frac{1}{2}}(1+\gamma_5\, \slash{s}_t)(1+\gamma_5)\right) 
\nonumber \\
&= 128 \, (p_{b} \cdot p_{\nu})\ (\bar{p}_t \cdot p_{\ell}) \, ,
\end{align}
where we have used the abbreviation
\begin{equation}
\label{ptbar}
\bar{p}_t^\mu=p_t^\mu - m_t s_t^\mu \, ,
\end{equation}
with $s_{t}$ denoting the polarization four-vector of the top quark. 
The scalar products in Eq.~(\ref{trace}) can be evaluated using explicit
representations of the pertinent four-vectors in the rest frame of the top quark.
From Fig.~\ref{Cazi} one has ($x=2E_{\ell}/m_t;\, y^2=m_W^2/m_t^2$)
\begin{eqnarray}
\label{momenta}
p_t &=& m_t(1;0,0,0)\,,\\
p_{\ell} &=& \frac{m_t}{2}x(1;0,0,1) \,,\nonumber \\
p_\nu &=&\frac{m_t}{2}(1-x+y^2)(1;-\sin\theta_\nu,0,\cos\theta_\nu) \,,\nonumber \\
p_b &=&\frac{m_t}{2}(1-y^2)(1;\sin\theta_b,0,\cos\theta_b) \,,\nonumber  \\
s_t &=&(0;\vec{P}_{t})= (0;\sin\theta_P \cos\phi,\sin\theta_P \sin\phi,\cos\theta_P) \,,\nonumber
\end{eqnarray}
where 
\begin{eqnarray}
\cos\theta_\nu &=& \frac{x(1-x+y^2)-2y^2}{x(1-x+y^2)} \,, \\
\cos\theta_b &=& \frac{2y^2-x(1+y^2)}{x(1-y^2)} \,.\nonumber
\end{eqnarray}
For the spin summed
squared matrix element we then obtain
\begin{equation}
\sum_{s_{b},s_{\ell},s_{\nu}}|M|^2 = 32 \, m_t^4 \, x(1-x)(1+\cos\theta_P)\, ,
\end{equation}
which, in the narrow width approximation for the $W$, leads to the partial rate
formulas Eqs.~(\ref{sm-bornterm1}) and (\ref{sm-bornterm2}).

The above derivation shows that the LO result $\Gamma \sim (1+\cos\theta_P)$ 
does not depend on the mass of the bottom quark. It does, however, depend
on the mass of the lepton. The lepton mass effect can be easily calculated
using the trace formula in Eq.~(\ref{trace}). One obtains
$|M|^{2} \sim 1+(1- \frac{1}{2}m_{\ell}^2/E_{\ell}^2+...)\cos\theta_P $.
The lepton mass correction is thus negligibly small since, in the narrow
resonance approximation for the $W^+$, the minimal lepton
energy is given by $E_{\ell}^{{\rm min}}=(m_W^4 + m_t^2m_{\ell}^2)/(2m_tm_W^2)$ and is thus very
much larger than the lepton mass appearing in the lepton mass correction.

Returning to the original current-current form (\ref{fierz1}) and its
Fierz--transformed form (\ref{fierz3}) it is clear that there will be no
azimuthal correlation, i.e. one has $\Gamma_C=0$ at the Born term level. 
It is nevertheless instructive and interesting to go through the exercise
to show that $\Gamma_C=0$ directly follows from 
$\Gamma_A=\Gamma_B$ if the rate is to remain positive definite over all
of phase space. We use a short--hand notation and write $A$ for 
$d\Gamma_A/dx$ and $B$ for $d\Gamma_B/dx$ etc.. 
With $A=B$ the angular decay distribution is given by (we set P=1)   
\begin{equation}
\label{ang-dist1}
\Gamma \sim A(1+\cos\theta_P + \frac{C}{A} \sin\theta_P \cos\phi)\,.
\end{equation}  
From the structure of Eq.~(\ref{ang-dist1}) one can immediately conclude that
the ratio $C/A$ necessarily has to vanish if the rate is to remain
positive definite over all of angular phase space. This can be seen in the
following way. Assume first that $C/A$ is positive. Set $\cos\phi=-1$ 
and expand the resulting decay distribution around
$\theta_P=\pi$ ($\theta_P\leqslant\pi$). One obtains
\begin{equation}
\label{ang-dist2}
\Gamma \sim A (\pi-\theta_P)(\frac{\pi-\theta_P}{2} - \frac{C}{A})\,.
\end{equation}
For any given value of $C/A$ the piece $(\pi-\theta_P)/2$ can
always be chosen small enough to render the rate to become negative.
If $C/A$ is assumed to be negative one chooses $\cos\phi=+1$ and goes 
through the same steps of arguments as before. 
The upshot is that $C$ has to be zero if one has $A=B$ in order for the rate 
to be positive definite everywhere. As mentioned before the explicit 
calculation using the form (\ref{fierz1}) or more directly (\ref{fierz3}) 
of course confirms this conclusion.   

\section{QCD NLO contribution to the azimuthal correlation  function $\Gamma_C$}
 
The ingredients for the NLO calculation are the virtual one-loop contributions
on the one hand, and the tree graph (real emission) contribution on the other hand. Both of 
these have been calculated before (as e.g. in \cite{fgkm02}) and we can make 
use of the previous results.

The virtual one-loop amplitudes are defined by covariant
 expansions ($ J^V_\mu = \bar{\psi}(b) \gamma_\mu \psi(t)$, 
$J^A_\mu =\bar{\psi}(b) \gamma_\mu \gamma_5 \psi(t) $):  

\begin{align}  
 \label{formfactor} 
 \langle b(p_b) | J^V_{\mu} | t(p_t) \rangle & \!= 
  \bar{u}(b) \left( \gamma_{\mu} F_1^V + p_{t, \mu} F_2^V +
  p_{b,\mu} F_3^V \right) u(t), \\ 
  \langle b(p_b) |J^A_{\mu} | t(p_t) \rangle & = 
  \bar{u}(b) \left( \gamma_{\mu} F_1^A + p_{t, \mu} F_2^A +
  p_{b,\mu} F_3^A \right) \gamma_5\, u(t).
 \end{align} 

 \noindent The Standard Model current combination is given by
 $ J^V_\mu - J^A_\mu $. At the one-loop level the form factors are
\cite{fgkm02} ($C_{F}=4/3$)
\begin{align}
  F_1^V & =  F_1^A = 1 \!-\! \frac{\alpha_s}{4 \pi} C_F \Big[
  4 \!+\! \frac{1}{y^2} \ln(1 \!-\! y^2)+
  2 \ln \Big( \frac{\Lambda^2}{\epsilon} \frac{1}{1 \!-\! y^2} \Big)
  \ln \Big( \frac{\epsilon}{1 \!-\! y^2} \Big) 
  \!+\! \nonumber\\ 
&\qquad\qquad  \!+\!\ln \Big( \frac{\epsilon}{1 \!-\! y^2} \frac{\Lambda^4}{(1 \!-\! y^2)^2} \Big)
  \!+\!  2 \mbox{Li}_2(y^2) \Big], \\
  F_2^V & =  - F_2^A = \frac{1}{m_t} \frac{\alpha_s}{4 \pi} C_F \,
  \frac{2}{y^2} \Big[+ 1 + \frac{1 - y^2}{y^2} \ln (1 - y^2) \Big], \\
  F_3^V & =  - F_3^A = \frac{1}{m_t} \frac{\alpha_s}{4 \pi} C_F \,
  \frac{2}{y^2} \Big[- 1 + \frac{2 \, y^2 \!-\! 1}{y^2} \ln (1 \!-\! y^2) 
  \Big],
 \end{align}
\noindent where a gluon mass regulator was used to regularize the gluon IR 
singularity. The scaled gluon mass and the scaled bottom quark mass are denoted by
 $ \Lambda = m_g / m_t $ and $\epsilon=m_{b}/m_{t}$.
 As mentioned earlier on, the logarithmic terms
in the gluon mass will not contribute to the azimuthal correlation function
and can therefore be dropped. The dilog function $ \mbox{Li}_2(x) $ is 
defined by
 \begin{equation}
  \mbox{Li}_2(x) := - \int\limits_{0}^{x} \frac{\ln(1 - z)}{z} \, dz\,.
 \end{equation} 
The tree graph contribution results from the square of the real gluon emission
graphs. For the corresponding hadron tensor one obtains~\cite{fgkm02} 
 \begin{align} 
 \label{Hadrontensor}
  {\cal H}^{\mu \nu} & = 4 \pi \alpha_s \, C_F \,
  \frac{4}{(k \!\cdot\! p_t)(k \!\cdot\! p_b)} \bigg\{ \nonumber \\
&  
  -i\frac{k \!\cdot\! p_t}{k \!\cdot\! p_b}   
  \Big(
  \epsilon^{\alpha \beta \mu \nu} \, (p_b \!-\! k) \!\cdot\! \bar{p}_t -
  \epsilon^{\alpha \beta \gamma \nu} (p_b \!-\! k)^{\mu} \, \bar{p}_{t,\gamma} + 
  \epsilon^{\alpha \beta \gamma \mu} (p_b \!-\! k)^{\nu} \, \bar{p}_{t,\gamma} 
  \Big) k_{\alpha} \, p_{b,\beta} 
  \nonumber \\ 
&  +  \frac{k \!\cdot\! p_b}{k \!\cdot\! p_t} \Big[ (\bar{p}_t \!\cdot\! p_t)
  \Big( k^{\mu} \, p_b^{\nu} + k^{\nu} \, p_b^{\mu} -
  k \!\cdot\! p_b \, g^{\mu \nu} -
  i \, \epsilon^{\alpha \beta \mu \nu} k_{\alpha} \, p_{b,\beta} \Big) 
  \nonumber \\ 
&-  (\bar{p}_t \!\cdot\! k) \Big( (p_t \!-\! k)^{\mu} \, p_b^{\nu} +
  (p_t \!-\! k)^{\nu} \, p_b^{\mu} - (p_t \!-\! k) \!\cdot\! p_b \, g^{\mu \nu}  -
  i  \epsilon^{\alpha \beta \mu \nu}
  (p_t \!-\! k)_{\alpha} p_{b,\beta} \Big) \Big]\nonumber \\
&   -  (\bar{p}_t \!\cdot\! p_b) \Big( k^{\mu} \, p_b^{\nu} +
  k^{\nu} \, p_b^{\mu} - k \!\cdot\! p_b \, g^{\mu \nu} -
  i  \epsilon^{\alpha \beta \mu \nu} k_{\alpha} \, p_{b,\beta} \Big) +
  (p_t \!\cdot\! p_b) \Big( k^{\mu} \, \bar{p}_t^{\nu} +
  k^{\nu} \, \bar{p}_t^{\mu} - k \!\cdot\! \bar{p}_t \, g^{\mu \nu} \Big) 
  \nonumber \\ 
&  -  (k \!\cdot\! p_b) \Big( p_t^{\mu} \, \bar{p}_t^{\nu} +
  p_t^{\nu} \, \bar{p}_t^{\mu} - p_t \!\cdot\! \bar{p}_t \, g^{\mu \nu} \Big) +
  (k \!\cdot\! p_t) \Big( (p_b \!+\! k)^{\mu} \, \bar{p}_t^{\nu} \!+\!
  (p_b \!+\! k)^{\nu} \, \bar{p}_t^{\mu} \!-\! (p_b \!+\! k) \!\cdot\! \bar{p}_t
  \, g^{\mu \nu} \Big) \nonumber \\
&  + 2 (k \!\cdot\! \bar{p}_t) p_b^{\mu} \, p_b^{\nu}  +
  i \Big(\epsilon^{\alpha \beta \mu \nu}  (k \!\cdot\! \bar{p}_t) +
  \epsilon^{\alpha \beta \gamma \mu}  k^{\nu} \bar{p}_{t,\gamma} -
  \epsilon^{\alpha \beta \gamma \nu}  k^{\mu} \bar{p}_{t,\gamma} \Big)
  p_{b,\alpha} \, p_{t,\beta} \nonumber \\
& +  i \Big(\epsilon^{\alpha \beta \mu \nu} \, (p_t \!\cdot\! \bar{p}_t) +
  \epsilon^{\alpha \beta \gamma \mu} \, p^{\nu}_t \bar{p}_{t,\gamma} -
  \epsilon^{\alpha \beta \gamma \nu} \, p^{\mu}_t \bar{p}_{t,\gamma} \Big)
  k_{\alpha} \, p_{b,\beta} \bigg\}
  + B^{\mu \nu} \cdot \Delta_{SGF} \,,
 \end{align}
where $k$ is the gluon momentum. The abbreviation $\bar{p}_t^\mu$ is defined in (\ref{ptbar}).
The hadron tensor has been written such that the IR singular part in the
hadronic tensor has been isolated in the term 
$B^{\mu \nu} \cdot \Delta_{SGF}$ where $B^{\mu \nu}$ is the Born term hadron
tensor. The remaining pieces are IR finite. 
Again, when calculating the azimuthal
correlation, the IR divergent term will not contribute and can thus be 
dropped. Its explicit form does not need to concern us here. 
It can be found in \cite{fgkm02}.   

In the following we will concentrate on the azimuthal correlation function.
For the fully differential azimuthal correlation function $d\Gamma_C/dx dz$ 
we find $(z=\frac{(p_b+k)^{2}}{m_{t}^{2}})$
\begin{align}
\frac{d\Gamma_C}{dx dz} &= \Gamma_F 2 \pi  \frac{m_W}{\Gamma_W}
C_F (-\frac{\alpha_s}{2\pi}) 6y^2  \left( M_{t}^C(x,z) +M_{\ell}^C(x)\delta(z)
\right),
\end{align}
where $M_{t}^C(x,z)$ and $M_{\ell}^C(x)$ denote the tree graph
and the virtual one--loop contribution, respectively. 
The virtual one--loop contribution is multiplied by
$\delta(z)$ since there is no gluon emission in the one--loop
contribution and hence one has $z=0$.

For the virtual one--loop contribution one finds
\begin{equation}
M_{\ell}^C(x)= -\sqrt{y^2(1-x)(x-y^2)}\bigg( \frac{1-x}{y^2} \bigg) \ln (1-y^2) \,.
\end{equation}
Integrating the one--loop contribution over the scaled lepton energy
one obtains
\begin{equation}
\int_{y^2}^1 dx \,M_{\ell}^C(x)=-\frac{\pi}{16}\frac{(1-y^2)^{3}}{y}\ln(1-y^2)\,.
\label{loopC} 
\end{equation}

The tree--graph contribution is rather more involved. One finds
\begin{align}
M_{t}^C(x,z)&=-\sqrt{y^2(1-x)(x-y^2)-xy^2z}\Big[\frac{y^2}{x\lambda^3}j_1
+ \frac{1}{\lambda^3}j_2 + \frac{x}{\lambda^3}j_3 \nonumber \\
&\qquad
+ 4Y_{p} \,\left( \frac{6y^2z}{x\lambda^{7/2}}j_4+\frac{1}{\lambda^{7/2}}j_5
+ \frac{x}{\lambda^{7/2}}j_6 \right) \, \Big] \, ,
\end{align}
where
\begin{align}
j_1 &= (1-y^2)^4 +(1-y^2)^{2}(25+2y^2-y^4)z  -4(1-y^2)(11+y^2+y^4)z^2  \nonumber \\
    & \quad +2(4-8y^2-3y^4)z^3+ (11+4y^2)z^4 -z^5 \, , \\
j_2 &= -(1-y^2)^4 (11+2y^2) + 2(1-y^2)^3(13-2y^2)z -4(3+2y^2+3y^4)z^2 \nonumber \\
    & \quad -2(5-23y^2+2y^4)z^3 +(7+2y^2)z^4 \,,\\
j_3 &= 12(1-y^2)^4 - 2(1-y^2)^2(6+7y^2)z -4(3-13y^2+2y^4)z^2 +2(6+5y^2)z^3\,,   \\
j_4 &= - (1-y^2)^3 -y^2(1-y^2)z +3z^2 -2z^3 \,, \\
j_5 &= 2(1-y^2)^5 + 5y^2(1-y^2)^3z   -(1-y^2)(11-23y^2+4y^4)z^2   \nonumber \\
    & \quad + (13-19y^2+14y^4)z^3-(3+10y^2)z^4 - z^5 \,, \\
j_6 &= -2(1-y^2)^5 -(1-y^2)^3(4-5y^2)z  +(1-y^2)(12-11y^2-3y^4)z^2 \nonumber \\ 
    &\quad -(4+15y^2-y^4)z^3- (2+y^2)z^4 \,,
\end{align}
with $ \lambda:=\lambda(1,y^2,z)=1+y^4+z^{2}-2(y^2+z+y^2 z)$ and 
$Y_{p}=\frac{1}{2}\ln \frac{1-y^2+z+\sqrt{\lambda}}{1-y^2+z-\sqrt{\lambda}}$.

In order to obtain the lepton energy spectrum $d\Gamma_C/dx$ the tree 
graph contribution has to be integrated over $z$ in the interval
$0 \leqslant z \leqslant (1-x)(1-\frac{y^2}{x})$. We have not been able to do this 
integration in closed form so the integration was done numerically. 

\begin{figure}[tbhp]
\begin{center}
\includegraphics[width=9cm]{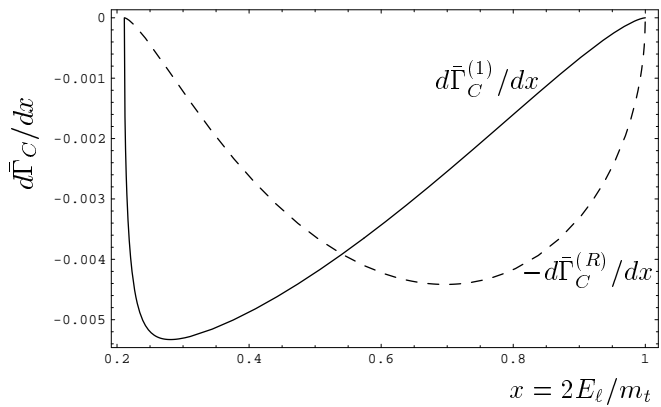}
\end{center}
\caption{\small \sf Lepton energy dependence of the azimuthal correlation functions: 
the  SM $O(\alpha_s)$ contribution $ d\Gamma_{C}^{(1)}/dx$ (solid line), 
and the right--chiral contribution $ d\Gamma_{C}^{(R)}/dx$ (dashed line).
With the bar we denote that they are scaled to the LO unpolarized
 total rate $ \Gamma_{A}^{(0)}$, i.e. 
$ d{\bar \Gamma}_{C}/dx =(\Gamma_{A}^{(0)})^{-1}d\Gamma_{C}/dx $. }
\label{spectrumC}
\end{figure}

We show the $x$--spectrum $d\Gamma_C/dx$ ($y^2 \leqslant x \leqslant 1$) in Fig.~\ref{spectrumC}
by adding the tree graph and virtual one--loop contributions $M_{t}^C(x)$ and $M_{\ell}^C(x)$. 
The azimuthal correlation function is small and negative over the whole spectrum and 
peaks at the lower end of the spectrum. The smallness of the azimuthal correlation 
function can be assessed by comparing the integrated azimuthal correlation function
with the integrated unpolarized Born term rate as done in Eq.~(\ref{numbers}). 
Fig.~\ref{spectrumC} also shows the spectrum of a possible
right--chiral contribution to the $t \to b$ transition which will be discussed in
the next section. Note, though, that the spectrum of the right--chiral contribution
is positive.

As a last point we calculate the fully integrated NLO azimuthal correlation 
function $\Gamma_C$. It turns out that the full analytical integration of the
tree graph contribution can be done by reversing the order of integrations,
i.e. by first integrating over $x$ in the limits $w_{-}\leqslant x \leqslant w_{+}$ where
\begin{equation}
w_\pm = \frac{1}{2}\left(1+y^2-z \pm \sqrt{\lambda(1,y^2,z)}\,\right)
\end{equation}
and then
over $z$ $(0 \leqslant z \leqslant (1-y)^2)$.  
We then obtain
\begin{align}
\label{treeC}
\int_0^{(1-y)^2}dz \int_{w_-}^{w_+}dx\, M_{t}^C(x,z) &=  \frac{\pi}{16} \Big\{
4y(4+3y^2-3y^4)\mbox{Li}_2(-y)\nonumber \\
&\quad -2(1-y^2)(8-7y+8y^2-5y^3)\ln(1+y) \nonumber  \\
&\quad +\frac{1}{3}y\left[6(1-y)^2(1-y-2y^2)+\pi^2(4+3y^2-3y^4)\right] 
                        \Big\}.
\end{align}

Finally we add up (\ref{loopC}) and (\ref{treeC}) to obtain
the NLO fully integrated azimuthal correlation function $\Gamma_C$. We find
\begin{align}
\Gamma_C^{(1)}&=\Gamma_F \, 2\pi \frac{m_W}{\Gamma_W} C_F (-\frac{\alpha_s}{2\pi})
\frac{3}{8}\pi y^2 \bigg\{ 4y(4+3y^2-3y^4)\mbox{Li}_2(-y)\nonumber\\
&\qquad -2(1-y^2)(8-7y+8y^2-5y^3)\ln(1+y) -\frac{(1-y^2)^{3}}{y}\ln(1-y^2)\nonumber \\ 
&\qquad +\frac{1}{3}y\big[6(1-y)^2(1-y-2y^2) +\pi^2(4+3y^2-3y^4)\big] \bigg\}\,. 
\end{align}

In the last step we combine our results for the azimuthal correlation rate with the
results for the unpolarized rate and the polar correlation rate from \cite{{cj94}}.
Numerically we obtain ($\Gamma^{{\rm NLO}} = \Gamma^{(0)} + \Gamma^{(1)}$\,\,;\,  
$\alpha_s(m_t^2)=0.107, y^2=0.211$)
\begin{align}
\label{numbers}
\frac{d\Gamma^{{\rm NLO}}}{d \cos\theta_P \, d\phi} &=
\frac{\Gamma_A^{(0)}}{4\pi}\Big[(1-8.54\%)+ (1-8.71\%) P \cos\theta_P  - 0.24\%\,P \sin\theta_P \cos\phi \Big] \\
\label{numbers2}
&=\frac{\Gamma_A^{{\rm NLO}}}{4\pi}\Big[1 + 0.998 P \cos\theta_P - 0.0026
\,P \sin\theta_P \cos\phi \Big] .
\end{align}

The radiative corrections to the rate $\Gamma_A$ and the polar correlation
function $\Gamma_B$ go in the same direction and are very close in magnitude.
The polar analyzing power therefore remains largely unchanged by the radiative
corrections as (\ref{numbers2}) shows  (100\% $\to$ 99.8\%). 
The azimuthal correlation generated by the
radiative corrections is quite small. It is safe to say that, if top quark decays
reveal a violation of the SM $(V-A)$ current structure in the azimuthal
correlation function which exceeds the 1\% level, the violation must have a 
non--SM origin.  

As discussed in Sec.~3 for the Born term contribution the positivity of the 
rate is an issue. We find that the NLO numerical rate values also satisfies 
positivity. Note that the positivity is not automatic
 in a NLO calculations. Although the NLO tree graph contribution is positive definite
the one-loop contributions is not necessarily positive 
since it involves an interference with the Born term amplitude.
To prove positivity we use a standard trigonometric identity to rewrite 
the NLO result (\ref{numbers}) as (we set $P=1$)
\begin{eqnarray}
&&\hspace{-0.7cm}\frac{\Gamma_A^{(0)}}{4\pi}\Big[ \Big(1+ \frac{\Gamma_A^{(1)}}{\Gamma_A^{(0)}}\Big)
+\Big(1+ \frac{\Gamma_B^{(1)}}{\Gamma_A^{(0)}}\Big)\! \cos\theta_P
+\frac{\Gamma_C^{(1)}}{\Gamma_A^{(0)}} \sin \theta_P \cos \phi \Big]\nonumber  \\ 
&=&\frac{\Gamma_A^{(0)}}{4\pi}\Big[\Big(1+ \frac{\Gamma_A^{(1)}}{\Gamma_A^{(0)}}\Big) 
+\sqrt{\Big(1+ \frac{\Gamma_B^{(1)}}{\Gamma_A^{(0)}}\Big)^2 
+\Big(\frac{\Gamma_C^{(1)}}{\Gamma_A^{(0)}} \cos \phi \Big)^2 } 
\sin (\theta_P + \delta) \Big] \, ,
\end{eqnarray}
where 
\begin{equation}
\tan \delta = \frac{\Gamma_A^{(0)}+\Gamma_B^{(1)}}{\Gamma_C^{(1)}\cos\phi} \, .
\end{equation}
For $\sin (\theta_P + \delta)=-1$ and $\cos\phi=\pm1$ the rate becomes minimal.
With the numbers in Eq.~(\ref{numbers}) one can check that the minimal value of the
rate is positive.

\section{Non-SM right-chiral quark current}


In order to be able to assess the size of the NLO contribution to the
azimuthal correlation we add a right-chiral piece to the quark current

\begin{eqnarray}
\bar{\psi}(b)\big[\gamma^\mu(1-\gamma_5)\big]\psi(t) \to
\bar{\psi}(b)\big[\gamma^\mu(1-\gamma_5)+ \delta_R\gamma^\mu
(1+\gamma_5)\big]\psi(t)
\end{eqnarray}
where $\delta_R$ parametrizes the strength of the right--chiral contribution.
From the discussion in Sec.~3 we anticipate that the right-chiral quark 
current will generate a nonvanishing azimuthal correlation. The current--current
matrix element involving the new right-chiral quark current will then read
\begin{align}
\label{right-chiral1}
M &= \delta_R \ \bar{u}(b)\gamma^\mu(1+\gamma_5)u(t)\,\,
\bar{u}(\nu)\gamma_\mu(1-\gamma_5)v(\ell) \\
\label{right-chiral2}
&= 
2 \,\delta_R \, \bar{u}(b)
(1-\gamma_5)v(\ell)  \, \bar{u}(\nu)(1+\gamma_5)u(t)\, 
\end{align}
where we have used a Fierz identity of the first kind to simplify
the matrix element (\ref{right-chiral1}). 

There are some indirect model dependent constraints on the strength of the 
right-chiral quark  $\delta_R \leqslant 0.004$ from an analysis of the rare decay 
$b \to s \gamma$ \cite{Fujikawa:1993zu,Larios:1999au,Burdman:1999fw}. In this
paper we take a phenomenological point of view and leave the size of $\delta_R$ 
unconstrained.

To start with we assume that $m_b=0$ or more generally 
$\delta_R \gg m_b/m_t$. For $m_b=0$ there will be no 
interference contribution from the interference of the left-- and 
right--chiral quark currents when squaring the full matrix element. The case
$\delta_R \simeq m_b/m_t$ will be discussed at the end of this section. Using the 
form (\ref{right-chiral2}) it is not difficult to obtain the
square of the right--chiral matrix element. One has 
\begin{align}
\!\!\sum_{s_{b},s_{\ell},s_{\nu}}|M|^2 &= 4\delta_R^{2} \, \rm{Tr} \Big( \slash{p}_\nu(1+\gamma_5)
\,\,(\slash{p}_t+m_t)\frac{1}{2}(1+\gamma_5\,\slash{s}_t)
(1-\gamma_5) \Big) \, \rm{Tr} \Big( \slash{p}_b(1-\gamma_5)\slash{p}_{\ell}
(1+\gamma_5) \Big)\nonumber \\ 
\label{m-squared2}
&=128\,\delta_R^2\,(p_\nu \cdot p_t+m_t\,p_\nu \cdot s_t)(p_b \cdot p_{\ell})\,.
\end{align}

The scalar products in Eq.~(\ref{m-squared2}) can again be evaluated using the
explicit representations of the pertinent four-vectors given in Eq.~(\ref{momenta}). 
Using again the narrow resonance approximation one has ($x=2E_{\ell}/m_t;\, y^2=m_W^2/m_t^2$)
\begin{align}
\frac{d\Gamma_A^{(R)}}{dx}&=\delta_R^2 \Gamma_F \, 2\pi\frac{m_W}{\Gamma_W} 6y^2(x-y^2)
(1-x+y^2)\, , \\
\frac{d\Gamma_B^{(R)}}{dx}&=\delta_R^2\Gamma_F \, 2\pi\frac{m_W}{\Gamma_W} 
\frac{6y^2(x-y^2)}{x}(2y^2-x(1+y^2-x)),  \\
\frac{d\Gamma_C^{(R)}}{dx}&= \delta_R^2\Gamma_F \, 2\pi\frac{m_W}{\Gamma_W} 
\frac{12y^2(x-y^2)}{x}\sqrt{y^2(1-x)(x-y^2)} . 
\end{align}
In Fig.~\ref{spectrumC} we show a plot of the spectrum of the azimuthal part of the right--chiral
contribution where we have fixed $\delta_R=0.051$ from arbitrarily setting
$|\Gamma_C^{(R)}|=|\Gamma_C^{(1)}|$, i.e. the two spectra in Fig.~\ref{spectrumC} have the same area.
One notes that, besides having a different sign, the $x$--dependence of the 
right--chiral contribution is harder
than that of the Standard Model. If the $x$--dependence can be measured it should not be 
difficult to differentiate between the two cases.

For the integrated rates we obtain ($y^2\leqslant x \leqslant  1$)
\begin{align}
\Gamma_A^{(R)}&=\delta_R^2\Gamma_F \, 2\pi\frac{m_W}{\Gamma_W}
y^2(1-3y^4+2y^6) \, ,  \\
\Gamma_B^{(R)}&=\delta_R^2\Gamma_F \, 2\pi\frac{m_W}{\Gamma_W} y^2(-1+12y^2-9y^4-2y^6+12y^4 \ln y^2)\, , \\
\Gamma_C^{(R)}&=\delta_R^2 \Gamma_F \, 2\pi\frac{m_W}{\Gamma_W}
\frac{3}{2}\pi y^3 (1-6y^2+8y^3-3y^4)  \, . 
\end{align}

Of course, for small values of $\delta_R$, i.e. when $\delta_R \simeq m_b/m_t$,
the interference between SM and non--SM--type contributions cannot be neglected. If
one takes a (one-loop) running $b$--quark mass of $m_b(m_t)= 1.79$ GeV and $m_t=175$ GeV
this would correspond to $\delta_R \simeq m_{b}/m_{t}= 0.0102$. 
If one takes a pole mass of $m_b=4.8$ GeV
this would correspond to $\delta_R \simeq m_{b}/m_{t}= 0.027$.
The contribution of the interference term to the differential rate is
\begin{align}
&\!\! \frac{d\Gamma^{(int)}}{dx}=-\delta_R \, \frac{m_{b}}{m_{t}} \Gamma_F 
\, 2\pi\frac{m_W}{\Gamma_W}  12y^2\Big[ y^2(1+P\cos\theta_P) +\sqrt{y^2(1-x)(x-y^2)}P\sin\theta_P \cos\phi
\Big].
\end{align}
Finally, integrating over $x$ one obtains
\begin{align}
&\Gamma^{(int)}= - \delta_R\, \frac{m_{b}}{m_{t}} \Gamma_F \, 2\pi\frac{m_W}{\Gamma_W} 
\frac{3}{2}y^3(1-y^2) \Big[ 8y(1+P\cos\theta_P) +\pi (1-y^2)P\sin\theta_P \cos\phi
\Big].
\end{align} 
It is curious to note that for $ \delta_R \simeq m_{b}/m_{t}$ the integrated 
interference and right--chiral contributions to $\Gamma_C$ tend to cancel each 
other, cf.
\begin{equation}
\Gamma_C^{(R)}+ \Gamma_C^{(int)}= \Gamma_A^{(0)} \textstyle{\frac{3}{2}}\pi \delta_R
(0.20\, \delta_R - 0.32\, \frac{m_{b}}{m_{t}})\,.
\end{equation}
If one takes $\delta_R \leqslant 0.004$ as suggested by the analysis of 
the rare decay $b \to s \gamma$ \cite{Fujikawa:1993zu,Larios:1999au,Burdman:1999fw}
one finds $|\Gamma_C^{(R)}+ \Gamma_C^{(int)}| \leqslant  4.7 \cdot 10^{-5}\, \Gamma_A^{(0)}$
using $m_{b}/m_{t}=0.0102$. This is far below the SM value 
$|\Gamma_C^{(1)}|= 2.4 \cdot 10^{-3}\, \Gamma_A^{(0)}$ that we have obtained in 
Eq.~(\ref{numbers}).

\section{Summary and conclusions}
We have calculated the $O(\alpha_s)$ corrections to an azimuthal
correlation observable in polarized top quark decay which vanishes at the Born 
term level. We have found that the $O(\alpha_s)$ corrections to this particular 
azimuthal correlation are quite small. If top quark decays
reveal a violation of the SM $(V-A)$ current structure in the azimuthal
correlation function which exceeds the 1\% level, the violation must have a 
non--SM origin.
  
We have used the helicity system for our analysis where the event plane lies in
the $(x,z)$--plane and the lepton momentum is along the $z$--axis. Other helicity
systems, where the $z$--axis is defined by the neutrino or the bottom quark jet,
provide independent probes of the polarized top quark decay dynamics. The Born
term angular correlations in these two additional helicity systems were studied in 
\cite{kp99}. The $O(\alpha_s)$ radiative corrections to the angular correlations 
in these helicity systems will be the subject of a future publication.

\end{document}